\documentclass[a4paper]{article}

\begin {document}

\title {\bf Relativistic Quantum Newton's Law For A Spinless Particle}
\author{A.~Bouda\footnote{Electronic address: 
{\tt bouda\_a@yahoo.fr}} \ and F.~Hammad\\
Laboratoire de Physique Th\'eorique, Universit\'e de B\'eja\"\i a,\\ 
Route Targa Ouazemour, 06000 B\'eja\"\i a, Algeria\\}

\date{\today}

\maketitle

\begin{abstract}
\noindent
For a one-dimensional stationary system, we derive a third order equation 
of motion representing a first integral  of the relativistic quantum 
Newton's law. We then integrate this equation in the constant 
potential case and calculate the time spent by a particle tunneling
through a potential barrier. 
\end{abstract}

\vskip\baselineskip



\newpage
\vskip0.5\baselineskip
\noindent
{\bf 1 \ \ Introduction }
\vskip0.5\baselineskip

The major obstacle in the quantization of gravity results from the fact that 
quantum mechanics in the context of Copenhagen interpretation is a 
probabilistic theory while general relativity describes gravity in a 
geometrical framework by linking the gravitational field to the curvature 
of space. In order to reconcile these two fundamental theories of the 
contemporary physics, a possible way consists first in obtaining a causal
and deterministic approach of quantum mechanics. In this spirit, Faraggi and 
Matone derived recently quantum mechanics from an equivalence postulate \cite 
{FM1,FM2} and showed by introducing a quantum transformation that the 
classical and the quantum potentials deform space geometry \cite 
{FM2,FM3}. This quantum transformation has allowed in Ref. \cite {BD1} to 
establish the quantum Newton's law for non-relativistic systems. The starting 
point is the quantum stationary Hamilton-Jacobi equation (QSHJE)
\begin {equation}
{1\over 2m} \left({\partial S_0 \over \partial x}\right)^2 + V(x)-E= 
{\hbar^2\over 4m}  \left[{3\over 2}\left( 
{\partial S_0 \over\partial x}\right)
^{- 2 }\left({\partial^2 S_0 \over \partial x^2}\right)^2-
\left( {\partial S_0 \over \partial x  }\right)^{- 1 }
\left({\partial^3 S_0 \over \partial x^3  }\right) \right] \; ,
\end {equation}
in which $S_0$, $E$ and $V$ are respectively the reduced action, the energy 
and the classical potential.
The solution of Eq. (1) is investigated in Refs. \cite{FM1,FM2,Fl1,Bo} .
It is shown in \cite{BD1} that it can be written as 
\begin {equation}
S_0=\hbar \ \arctan {\left ( a  {\phi_1 
\over \phi_2 } +b \right )} +\hbar \lambda \; ,
\end {equation}
where $(\phi_1,\phi_2)$ is a set of two real independent solutions
of the Schr\"odinger equation  
\begin {equation}
-{\hbar^2 \over 2m} {d^2 \phi \over dx^2 } + V(x) \phi = E \phi\; , 
\end {equation}
and \ $(a,b,\lambda)$ \ are real integration constants satisfying 
the condition $a \not= 0$. In Eq. (2), $S_0$ depends also on 
the energy $E$ through the solutions $\phi_1$ and $\phi_2$.
In contrast with Bohm's theory, it is shown in Refs.
\cite{FM1,FM2,Bo} that it is possible to relate the 
reduced action $S_0$ to the Schr\"odinger wave function in a 
unified form both for bound and unbound states so that
the conjugate momentum 
\begin{equation}
P =  {\partial S_0\over\partial x}
  =  {\hbar a W \over {\phi_2^2 + 
         (a\phi_1 + b\phi_2)^2}} \;  
\end {equation}
never has a vanishing value. In (4), $W=\phi_1'\phi_2-\phi_1\phi_2'$ is a 
constant representing the Wronskian of $(\phi_1,\phi_2)$.

By taking advantage of the fact that the 
solution of (1) is known, the 
fundamental relation
\begin {equation}
\dot{x}{\partial S_0 \over \partial x}=
2[E-V(x)] \; ,
\end {equation}
is derived \cite{BD1}. It is also showed that this last equation
leads to a third order differential equation representing the first 
integral of the quantum Newton's law (FIQNL)
\begin {eqnarray}
(E-V)^4-{m{\dot{x}}^2 \over 2}(E-V)^3+{{\hbar}^2 \over 8} 
{\left[{3 \over 2}
{\left({\ddot{x} \over \dot{x}}\right)}^2-{\dot{\ddot{x}} \over \dot{x}} 
\right]} (E-V)^2\hskip15mm&& \nonumber\\
-{{\hbar}^2\over 8}{\left[{\dot{x}}^2 
{d^2 V\over dx^2}+{\ddot{x}}{dV \over dx}
 \right]}(E-V)-{3{\hbar}^2\over 16}{\left[\dot{x}
{dV \over dx}\right]^2}=0 \; .
\end {eqnarray}
The solution $x(t)$ of this equation will contain the two 
usual integration constants $E$ and $x_0$ and two additional 
constants that we will call the non-classical integration 
constants. All these constants can be determined by 
the knowledge of $x(t_0)$, $\dot{x}(t_0)$, $\ddot{x}(t_0)$ 
and $\dot{\ddot{x}}(t_0)$.

In this paper, we will attempt to twin special relativity with 
quantum mechanics in the context of the trajectory representation. 
In Section 2, we extend the quantum law of motion obtained in 
\cite {BD1} to relativistic spinless systems. In Section 3, we 
apply our results in the constant potential case and derive the time 
delay in tunneling through a barrier potential. Section 4 is devoted 
to comments about the paradox arising from the twinning of 
relativity postulates and quantum theory.

\vskip0.5\baselineskip
\noindent
{\bf 2\ \ The relativistic quantum law of motion}
\vskip0.5\baselineskip

For a spinless particle, the relativistic quantum stationary 
Hamilton-Jacobi equation (RQSHJE) is given in one dimension by
\cite{FM2,BFM}

\begin {eqnarray}
{1\over 2m} \left({\partial S_0 \over \partial x}\right)^2 + 
{mc^2 \over 2 }-{1 \over 2mc^2 } [E-V(x)]^2 \hskip40mm&& \nonumber\\
= {\hbar^2\over 4m}  \left[{3\over 2}\left( 
{\partial S_0 \over\partial x}\right)
^{- 2 }\left({\partial^2 S_0 \over \partial x^2}\right)^2-
\left( {\partial S_0 \over \partial x  }\right)^{- 1 }
\left({\partial^3 S_0 \over \partial x^3  }\right) \right] \; .
\end {eqnarray}
In the context of the equivalence postulate \cite{FM1,FM2}, this 
equation leads to the stationary Klein-Gordon equation \cite{BFM}
\begin {equation}
-{\hbar^2\over2m} {d^2\phi\over {dx}^2}+{{m^2c^4-[E-V]^2}\over2mc^2} \phi=0\; .
\end {equation}
As in the non-relativistic case, one can check that the general solution of 
(7) can be written in the 
following form
\begin {equation}
S_0=\hbar\ \arctan \left(a{\phi_1\over\phi_2}+b\right)+ \hbar \lambda\; ,
\end {equation}
where $(\phi_1,\phi_2)$ is a set of two real independent solutions of 
the Klein-Gordon equation and $a$, $b$ and $\lambda$ 
are real parameters satisfying the condition $a \ne 0$.
With the same procedure developed in the non-relativistic case 
\cite{FM2,Bo}, we can start from the expression of $S_0$ to show 
that the Klein-Gordon wave function can be written in the unified 
form
$$
\phi(x)= \left({\partial S_0 \over \partial x }\right)^{-1/2}
\left[\alpha\ \exp\left({i\over\hbar }S_0\right)
+\beta\ \exp\left(-{i\over\hbar }S_0\right)\right]
$$
both for bound and unbound states, $\alpha$ and  $\beta$ being 
complex constants. This form of $\phi$ guarantees that the 
conjugate momentum, $P = \partial S_0 / \partial x $, never has 
a vanishing value.

In order to establish the equation of motion, let us appeal to the 
coordinate transformation introduced by Faraggi and Matone 
\cite{FM2,FM3} in the non-relativistic case,
$$
x \to \hat{x}\; ,
$$
after which we require that the RQSHJE takes the classical form
\begin {equation}
{1\over 2m} \left({\partial\hat{S_0} \over \partial\hat{x}}\right)^2 + 
{mc^2 \over 2 }-{1 \over 2mc^2 } [E-{\hat{V}(\hat{x})}]^2=0\; ,
\end {equation}
and $S_0(x)$ and $V(x)$ be invariant
$$
\hat{S_0}(\hat{x})=S_0(x),\ \ \ \ \ \ \ \ \ \ \hat{V}(\hat{x})=V(x).
$$
Then, Eq. (10) can be rewritten as
\begin {equation}
 {1\over 2m} {\left({\partial S_0 \over \partial x}\right)^2}
\left(\partial x \over \partial \hat{x} \right)^2 + 
{mc^2 \over 2 }-{1 \over 2mc^2 } [E-V(x)]^2=0\; .
\end {equation}
Substituting in this last equation $\partial S_0/ \partial x$ by $P$
and $E$ by $H$, we deduce that 
\begin {equation}
H(x,P)=\sqrt {P^2c^2 \left( \partial x \over\ \partial \hat{x}\right)^2+m^2c^4}
\; +V(x)\; ,
\end {equation}
which leads to the canonical equation
\begin {equation}
\dot {x}={\partial H\over\partial {P}}={Pc \left(\partial x/\partial \hat
 {x}\right)^2\over\ \sqrt {P^2 \left(\partial x/\partial \hat {x}\right)
^2+m^2c^2}}\; .
\end {equation}
We would like to indicate that in (12) and (13) the function 
$ \partial x /\partial \hat {x} $ depends implicitly on $E$. 
However, in order to guarantee that the quantum Hamiltonian (12) 
and the resulting canonical equation (13) will reproduce the classical 
results in the classical limit, we have not taken into account this 
dependence in the substitution of $E$ by $H$ in (11) . 
In fact, comparing (7) and (11) we see that 
$\partial x/\partial \hat {x}$ goes to 1 when $\hbar\to 0$. 
It follows that Eqs. (12) and (13)
respectively reproduce the well-known classical relations
\begin {equation}
(H-V)^2=P^2c^2+m^2c^4\; ,
\end {equation}
and
\begin {equation}
P={m\dot {x}\over\ \sqrt {1-{ \dot {x}}^2/c^2}}\; 
\end {equation}
for relativistic systems. Substituting in (13) $P$ by $\partial S_0/
 \partial x$ and then using (11), we get to the fundamental relation 
\begin {equation}
\dot {x}{ \partial S_0\over \partial x}=E-V(x)-{m^2c^4\over\ E-V(x)}\; .
\end {equation}
Firstly, in the classical limit, $\hbar\to 0$, since the conjugate momentum 
reduces to the expression given in (15), we can check that relation 
(16) reproduces the well-known classical relation of energy conservation
\begin {equation}
E={mc^2\over\ \sqrt {1-{\dot {x}}^2/c^2}}+V(x)
\end {equation}
for relativistic systems. Secondly, if we make in Eq. (16) the 
substitution
\begin {equation}
E^{nr}=E-mc^2\; ,
\end {equation}
where $E^{nr}$ represents the energy of the system without the 
energy at rest, and use the non-relativistic approximation 
$E^{nr}-V(x)\ll mc^2$, we can check that (16) 
reduces to the quantum relation (5) established in \cite {BD1} for 
non-relativistic systems.

For any potential $V(x)$, $\partial S_0/ \partial x$ can be
determined from Eq. (9). It follows that Eq. (16) represents the
relativistic quantum equation of motion. It is a first order equation
and contains three integration constants $(E, a, b)$. The solution 
$x(t)$ will contain a further constant.
As in the non-relativistic case \cite{BD1}, all these constants can be
determined by the initial conditions
\[
x(t_0)=x_0,\ \ \ \ \ \  
 \dot {x}(t_0)=\dot {x}_0,\ \ \ \ \ \
\ddot {x}(t_0)=\ddot {x}_0, \ \ \ \ \ \ 
\dot{\ddot {x}}(t_0)=\dot{\ddot {x}}_0\; .
\]
Now, let us derive the first integral of the relativistic quantum
Newton's law (FIRQNL). For this purpose, let us use relation (16) 
to compute the derivatives
\begin {equation}
{\partial^2 S_0\over {\partial x}^2}=-
\left[1+{m^2c^4\over (E-V)^2}\right]{1\over \dot {x}}{dV\over dx}
+\left[-1+{m^2c^4\over (E-V)^2}\right](E-V){\ddot {x}\over 
{\dot x}^3}
\end {equation}
and
\begin {eqnarray}
{\partial^3 S_0\over{ \partial x}^3}=\left[1+{m^2c^4\over (E-V)^2}\right]
\left[2{\ddot {x}\over {\dot {x}}^3}{dV\over dx}-{1\over \dot {x}}{d^2V
\over {dx}^2}\right]\hskip40mm&& \nonumber\\
-{2m^2c^4\over (E-V)^3}{1\over \dot {x}}\left[{dV\over dx}\right]^2 
+(E-V)\left[-1+{m^2c^4\over (E-V)^2}\right]\left[{\dot {\ddot {x}}\over 
{\dot {x}}^4}-{3{\ddot {x}}^2\over {\dot {x}}^5}\right]\; . 
\end {eqnarray}
Substituting these expressions in the RQSHJE given by (7), we can obtain 
the FIRQNL
\begin {eqnarray}
[m^2c^4-(E-V)^2]^3\left[m^2c^4-
(E-V)^2\left(1-{{{\dot{x}}^2}\over {c^2}}\right)
\right] \hskip35mm\nonumber\\
+{{{\hbar}^2}\over 2}[m^2c^4-(E-V)^2]^2
[E-V]^2\left[{\frac {3}{2}}{{{\ddot{x}}^2}\over {{\dot {x}}^2}}-
{{\dot {\ddot {x}}}\over {\dot {x}}}\right] \hskip28mm\nonumber\\
+{{\hbar}^2\over 2}[m^4c^8-(E-V)^4][E-V]\left[{\dot {x}}^2{{d^2V}\over 
{{dx}^2}}+{\ddot{x}}{{dV}\over {dx}}\right] \hskip18mm\nonumber\\
+{{{\hbar}^2}\over 4}[m^4c^8-10m^2c^4(E-V)^2-3(E-V)^4]
\left[{\dot {x}}{dV\over dx}\right]^2=0\; ,
\end {eqnarray}
in which we see the presence of the energy $E$. Thus, Eq. (21) represents the
equation of energy conservation for relativistic quantum spinless systems.
As in the non-relativistic case \cite {BD1}, we can reproduce (21) from (16)
in which, by using (9), we express $\partial S_0/ \partial x$ in terms of the
independent solutions $\phi_1$ and $\phi_2$ of Klein-Gordon's equation.
In contrast with relation (16), in order to solve (21), one does not
need to use the Klein-Gordon equation. However, the two equations are
equivalent. Of course, relation (16) does not depend on the choice of
the couple of solutions $(\phi_1, \phi_2)$ of Klein-Gordon's equation. In 
fact,  let us consider another couple $(\theta_1, \theta_2)$. As in 
Ref. \cite {BD2}, we can check that it is possible to find two 
parameters $( \tilde {a}, \tilde {b})$, which we must use instead of 
$(a, b)$ in expression (9) of the reduced action, in such a way as to get 
the conjugate momentum $ \partial S_0/ \partial x$ invariant.

Now, let us examine the classical and the non-relativistic limits for 
Eq. (21). Firstly, remark that if we put $\hbar=0$ in (21) we 
reproduce the well-known classical relation (17) of energy conservation 
for relativistic systems. Secondly, in the non-relativistic 
approximation $E^{nr}-V(x)\ll mc^2$, if we make the substitution (18), 
we check that (21) reduces to the FIQNL given by (6) and established 
in \cite {BD1} for non-relativistic systems.

The last point we will examine concerns the possibility of reproducing 
the fundamental relation (16) by appealing to the quantum version of Jacobi's
theorem \cite{BD1}
\begin {equation}
t-t_0=\left[{\partial \hat{S}_0(\hat{x})
\over\partial E}\right]_{\hat{x}=cte} \; . 
\end {equation} 
Taking the derivative with respect to $\hat {x}$ and then using Eq. (10), we
get to
\begin {eqnarray}
{dt\over d\hat {x}} & = & {\partial\over \partial \hat {x}}{\partial \hat{S}_0
(\hat{x})\over\partial E}  \nonumber\\
 & = & {\partial\over \partial E}{\partial \hat{S}_0(\hat{x})\over\partial 
\hat{x}}  \nonumber\\
 & = & {\partial\over \partial E}\sqrt { {[\hat {V}(\hat {x})-E]^2\over c^2}
-m^2c^2} \; .\nonumber 
\end {eqnarray}
As $\hat {V}(\hat {x})=V(x)$, we deduce that
\begin {equation}
{dt\over dx}{\partial x\over \partial \hat {x}}={E-V(x)\over c\sqrt{[E-V(x)]^2
-m^2c^4}}\; .
\end {equation}
From Eq. (11), we write
\begin {equation}
{\partial x\over \partial \hat {x}}=\sqrt{{[E-V(x)]^2\over c^2}-m^2c^2}
\;{1\over \partial S_0/\partial x}\; .\nonumber
\end {equation}
Substituting this last expression for $\partial x/\partial \hat {x}$ in (23),
we get to relation (16) from which we have deduced the FIRQNL, Eq. (21).

\vskip0.5\baselineskip
\noindent
{\bf 3\ \ The constant potential case}
\vskip0.5\baselineskip

Let us now consider the case for which the potential is constant $V(x)=V_0$
and set
\begin {equation}
{\epsilon}=E-V_0\; .
\end {equation}
The fundamental relation (16) takes the form
\begin {equation}
{\partial S_0 \over \partial x}dx=
          {\epsilon^2 -m^2c^4 \over \epsilon }dt\; .
\end {equation}
Let us begin with the case for which ${\epsilon}^2> m^2c^4$. Using
expression (9) for $S_0$ and choosing 
\begin {equation}
\phi_1 = \sin{\left({\sqrt{\epsilon^2-m^2c^4} \over \hbar c}x \right)}, 
\ \ \ \ \ \ \ \\ \ \ \ 
\phi_2 = \cos{\left({\sqrt{\epsilon^2-m^2c^4} \over \hbar c}x \right)}
\end {equation}
as independent solutions of Klein-Gordon's equation, relation (26) 
leads to 
\begin {equation}
x(t)={\hbar c\over \sqrt{\epsilon^2-m^2c^4}}\;
     \arctan\left[{1 \over a}\tan\left({\epsilon^2-m^2c^4\over 
     \hbar\epsilon }\;(t-t_0)\right)-{b \over a}\right]\; .
\end {equation}
The constant $\lambda$ appearing in (9) is absorbed in the 
term containing the integration constant $t_0$.
Note that for $a=1$ and $b=0$, Eq. (28) reduces to the classical one
\begin {equation}
{x(t)}={{{\sqrt{{\epsilon^2-m^2c^4}}\over {\epsilon}}}\;c(t-t_0)}
\end {equation}
for relativistic systems moving in a constant potential.

As in the non-relativistic case \cite{BD2}, the arctangent function 
is contained between $-\pi/2$ and $\pi/2$ and, therefore, it is necessary 
to add in the right hand side of (28) a constant which must be readjusted 
after every interval of time in which the tangent function goes from 
$-\infty $ to $+\infty $. Then, the continuity of 
$x(t)$ is guaranteed by rewriting (28) as
\begin {equation}
x(t)={\hbar c \over \sqrt{{\epsilon^2-m^2c^4}}}\;
     \arctan\left[{1 \over a}\tan \left({{\epsilon^2-m^2c^4} \over 
     {\hbar{\epsilon}}}\;(t-t_0)\right) - {b \over a }\right]+{\pi 
     {\hbar}c\over {\sqrt{{\epsilon^2-m^2c^4}}}}\;n
\end {equation}
with
\begin {equation}
t \in {\left[ t_0 + {\pi \hbar \epsilon \over \epsilon^2-m^2c^4}
\left(n-{1 \over 2} \right), t_0 + {\pi \hbar \epsilon\over \epsilon^2-m^2c^4}
\left(n+\frac{1}{2} \right)\right]}\; , 
\end {equation}
for every integer number $n$.

As in the non-relativistic case \cite{BD2}, if we choose the initial  
conditions in such a way as to have $t_0=0$, in $(t, x)$ plane all the 
trajectories corresponding to different values of $a$ and $b$, even the 
classical one $(a=1, b=0)$, pass through some points constituting nodes. 
The coordinates of these nodes are
$$
t_n={\pi \hbar \epsilon \over \epsilon^2-m^2c^4}
\left(n+{1 \over 2} \right),\ \ \ \ x(t_n)={\pi \hbar c \over {\sqrt{\epsilon^2
-m^2c^4}}}\left(n+{1 \over 2} \right)\; .
$$ 
We see that $x(t_n)$ is independent of $a$ and $b$. The distances between two 
adjacent nodes on the time axis
\begin {equation}
\Delta t_n=t_{n+1}-t_n={\pi \hbar \epsilon \over \epsilon^2-m^2c^4}\; , 
\end {equation}
and the space axis
\begin {equation}
\Delta x_n=x(t_{n+1})-x(t_n)={\pi \hbar c \over {\sqrt{\epsilon^2
-m^2c^4}}}\; ,
\end {equation}
are both proportional to $\hbar$. Therefore, in the classical limit, 
$\hbar\to 0$, the nodes become infinitely close. Furthermore, by 
examining the sign of the velocity, 
\begin {eqnarray}
\dot{x}(t)={ac\sqrt{\epsilon^2-m^2c^4}\over \epsilon}
           \left\lbrace a^2 \cos^2\left({\epsilon^2-m^2c^4\over \hbar 
           \epsilon} \;t \right)\right. \hskip30mm&& \nonumber\\
           +\left. \left[ \sin\left({\epsilon^2-m^2c^4\over \hbar 
           \epsilon}\;t\right)-b\ \cos\left({\epsilon^2-m^2c^4\over
           \hbar \epsilon}\;t\right)\right]^2\right\rbrace^{-1} \; , 
\end {eqnarray}
we see that the function $x(t)$ is increasing (decreasing) for 
$a\epsilon>0$ (for $a\epsilon<0$). With the same reasoning as the one 
of the non-relativistic case \cite {BD2}, we can deduce that in the 
classical limit, $\hbar\to 0$, all the quantum trajectories tend to 
be identical to their corresponding classical one. So, there is no 
residual indeterminacy \cite {Fl2}.

It is interesting to remark that for some values of $a$ and $b$, the 
instantaneous velocity can be higher than the light speed. However, 
this is not the case for the mean velocity between two adjacent nodes
\begin {equation}
v={\Delta x_n\over \Delta t_n}={c\sqrt{\epsilon^2-m^2c^4}\over \epsilon}
\; ,
\end {equation}
where we have used (32) and (33). Note that expression (35)  
represents the classical velocity and can also be reproduced 
from (34) by putting $a=1$ and $b=0$. 

Now, let us consider the case $\epsilon^2<m^2c^4$. Using expression 
(9) for $S_0$ and choosing 
\begin {equation}
\phi_1 = \exp{\left({\sqrt{m^2c^4-\epsilon^2} \over \hbar c}x \right)}, 
\ \ \ \ \ \ \ \ \ \ \ \ 
\phi_2 = \exp{\left(-{\sqrt{m^2c^4-\epsilon^2} \over \hbar c}x \right)}\; ,
\end {equation}
Eq. (26) leads to
\begin {equation}
x(t)={\hbar c\over 2\sqrt{m^2c^4-\epsilon^2}}\ln\left|{1\over a}\tan\left[
{m^2c^4-\epsilon^2\over \hbar \epsilon}(t-t_0)\right]+{b\over a}\right|\; ,
\end {equation}
where $t_0$ is a real integration constant. 
Eq. (37) represents the relativistic quantum time equation 
for a particle moving in a constant potential in the case where 
$\epsilon^2<m^2c^4$. The velocity is given by
\begin {equation}
\dot {x}(t)={c\over 2\epsilon}\sqrt{m^2c^4-\epsilon^2}\;{1+\tan^2\left[{
(m^2c^4-\epsilon^2)(t-t_0)/\hbar \epsilon}\right]\over b+\tan\left[{
(m^2c^4-\epsilon^2)(t-t_0)/\hbar \epsilon}\right]}\; .
\end {equation}
It is clear that if the particle enters the domain where $\epsilon^2<m^2c^4$
at the time $t_0$, its velocity becomes infinite at the times $t_1=t_0+\pi
\hbar\epsilon/2(m^2c^4-\epsilon^2)$ for $0\leq \epsilon<mc^2$ and $t_2=
t_0-\pi\hbar\epsilon/2(m^2c^4-\epsilon^2)$ for $-mc^2<\epsilon\leq 0$.

For example, in the case of a rectangular potential barrier
\[
V(x) = \left\{ \begin{array}{cc}
               0, & \  \  x<0 \\ [.1in]
               V_0, & \  \  0 \leq x \leq q \\ [.1in]
               0, & \  \ x>q \; ,
               \end{array} 
       \right.
\label{eq:ve}
\]
after we express $t$ in terms of $x$ from (26) with the use of (36), 
the time delay in tunneling through this barrier is
\begin {eqnarray}
T(q) \equiv t(q)-t(0) 
={\hbar\epsilon\over \epsilon^2-m^2c^4}\left\lbrace\arctan\left[a
\exp\left({2\sqrt{m^2c^4-\epsilon^2}\over \hbar c}\;q\right)+b\right]\right. 
\hskip4mm&&\nonumber\\
\left. -\arctan(a+b)\right\rbrace \; .
\end {eqnarray} 
Here, we have assumed that $a<0 \;$ if $\;0<\epsilon<mc^2 \;$ and  
$\; a>0 \;$ if $\; 0>\epsilon>-mc^2$. For a thin barrier 
$(\sqrt{m^2c^4 -\epsilon^2}\;q/\hbar c \- \ll 1)$ and a thick one 
$(\sqrt{m^2c^4-\epsilon^2}\;q
/\hbar c\gg 1)$, Eq. (39) turns out to be
\begin {equation}
T(q)=-{2a\epsilon \over c[1+(a+b)^2]\sqrt{m^2c^4-\epsilon^2}}\;q
\end {equation}
and
\begin {equation}
T(q)={\hbar\epsilon\over \epsilon^2-m^2c^4}\left[\pm {\pi\over 2}-
\arctan(a+b)\right]
\end {equation}
respectively. In (41), the signs $+$ and $-$ which precede 
$\pi/2$ correspond respectively to the cases $0>\epsilon>-mc^2$ 
and $0<\epsilon<mc^2$.  If we make the substitution 
$\epsilon^{nr}=\epsilon-mc^2$ and use the non-relativistic 
approximation, $\epsilon^{nr}\ll mc^2$, Eqs. (40) 
and (41) reproduce the results of Ref. \cite {BD2} obtained for 
non-relativistic systems. We mention that in different contexts, 
other authors \cite {Ha,Fle,Fl3} also investigated the problem of 
time delay in tunneling for non-relativistic systems.

\vskip0.5\baselineskip
\noindent
{\bf 4\ \ Discussion}
\vskip0.5\baselineskip

To conclude, we would like to tackle the conflict which seems to appear 
between quantum mechanics and special relativity. We remarked above that, 
in the case where $\epsilon^2>m^2c^4$ as well as in the case where 
$\epsilon^2<m^2c^2$, the instantaneous velocity can be higher than 
the light speed. In the classically allowed case $(\epsilon>mc^2)$, as 
in the non-relativistic case \cite {BD2}, the distance between two 
adjacent nodes is related to de Broglie's wavelength 
\begin {equation}
\lambda={2\pi\hbar\over P^{cl}}
\end {equation}
by
\begin {equation}
\Delta x_n={\lambda\over 2}.   
\end {equation}
In (42), $P^{cl}$ represents the classical conjugate momentum given in 
the right hand side of Eq. (15). Relation (43) is established with the 
use of (14), (25) and (33). We remarked that for any quantum 
trajectory, the mean velocity between two adjacent nodes is the same 
as the classical one. Then, the above result indicates that when we 
consider problems in which de Broglie's wavelength can be disregarded, 
the conflict between quantum mechanics and relativity postulates 
disappears. It is only on the microscopic scale, inside the intervals
separating adjacent nodes, that these postulates seem to be violated.
This conclusion reminds us the fact that in the standard quantum 
mechanics the energy conservation law can be violated for short 
durations. In the classically forbidden regions, there are 
no nodes, no classical limit and then, the velocity tends quickly to 
be infinite. 

We would like to add that even the non-relativistic case, for Floyd's 
formulation \cite {Fl2,Fl4} and for the one presented in 
Refs. \cite {BD1,BD2}, the velocity of a free particle is not constant.
Despite the presence of the quantum potential, which in our point of 
view is part of the kinetic term \cite {BD1}, the above remark indicates 
that the non-relativistic quantum mechanics, in the context of its 
trajectory interpretation, seems to be in conflict with Galilee's 
relativity principle. However, on the scale where de Broglie's 
wavelength can be disregarded, this conflict disappears.

\vskip\baselineskip
\noindent
{\bf REFERENCES}

\end {document}